\documentstyle[aps]{revtex}
\def\case#1/#2{\textstyle\frac{#1}{#2} }
\begin{document}
\begin{titlepage}
%%%%%%%%%%%%%%%%%%%%%%%%%%%%%%%%%%%%%%%%%%%%%%%%%%%%%%%%
\title{\bf Shrinking and the True Power Spectrum at Decoupling}
\author{
{\sc B. A. C. C. Bassett $^{1 \dagger}$ ,
P. K. S. Dunsby $^{1,2}$ and G. F. R. Ellis $^{1,3}$}\\
\normalsize{1. {\it Department of Applied
Mathematics, University of Cape Town,}}\\
\normalsize{{\it Rondebosch
7700, Cape Town, South Africa.}}\\
\normalsize{2. {\it School of
Mathematical Sciences, Queen Mary and Westfield College,}}\\
\normalsize{{\it University of London, Mile End Road, E1 4NS, UK.}}\\
\normalsize{3. {\it Scuola Internazionale Superiore di Studi Avanzati,}}\\
\normalsize{{\it Strada Costiera 11, 34014 Trieste,
Italy}} }
\date{$\mbox{}$ \vspace*{0.3truecm} \\ \today}
%%%%%%%%%%%%%%%%%%%%%%%%%%%%%%%%%%%%%%%%%%%%%%%%%%%%%%%%
\maketitle
\thispagestyle{empty}
%%%%%%%%%%%%%%%%%%%%%%%%%%%%%%%%%%%%%%%%%%%%%%%%%%%%%%%%
\begin{abstract}
%%%%%%%%%%%%%%%%%%%%%%%%%%%%%%%%%%%%%%%%%%%%%%%%%%%%%%%%
In this paper we examine the change in the estimated spatial power
spectra at decoupling due to the effects of our clumpy universe which
modify observational distances. We find that scales at decoupling can
be significantly underestimated in our approximation of
neglecting the shear of the ray bundle. We compare our results with other
work on lensing and speculate on the implications for structure formation.
In particular we examine a proposal to use the position of the first
Doppler peak to determine $\Omega$, and find that shrinking will
modify the esimated curvature, so that it must be included to obtain
an accurate estimate of $\Omega$. Finally we consider future applications
and improvements of our results.
%%%%%%%%%%%%%%%%%%%%%%%%%%%%%%%%%%%%%%%%%%%%%%%%%%%%%%%%
\end{abstract}
%%%%%%%%%%%%%%%%%%%%%%%%%%%%%%%%%%%%%%%%%%%%%%%%%%%%%%%%
\vspace*{0.2truecm}
\begin{center}
{\it Subject headings:\\ cosmology\,-\,gravitational lensing: structure
formation, cosmic microwave background }
\end{center}
\noindent
UCT preprint-94/4,\\
gr-qc/9405006,\\
To appear in the {\it Physical Review} {\bf D}\\
$\dagger$ email: bruce@maths.uct.ac.za
\end{titlepage}
%%%%%%%%%%%%%%%%%%%%%%%%%%%%%%%%%%%%%%%%%%%%%%%%%%%%%%%%
\section{Introduction} \label{sec: Intro}
%%%%%%%%%%%%%%%%%%%%%%%%%%%%%%%%%%%%%%%%%%%%%%%%%%%%%%%%
The present view of the evolution of structure in the universe is
that of growth of intially linear perturbations via gravitational
instability. If these perturbations are Gaussian and statistically
homogeneous and isotropic, then they are completely characterised by their
power spectra: the square of the Fourier transform amplitudes. In general
there may have been scalar (density), vector (vortex) and tensor (gravitational
wave) perturbations present at decoupling which would have left imprints on
the cosmic microwave background radiation (CMBR). The power spectra are
therefore fundamental entities in modern observational and theoretical
cosmology, linking the early, quantum universe with the present one, providing
a bridge between astronomy, cosmology and particle physics.

Recent interest in the CMBR and the power spectra at decoupling has again
reached a high point with the release of the 2nd year {\it COBE} DMR
temperature maps \cite{bi:Wright94}. The analysis of the data
\cite{bi:Smoot92,bi:Wright92} generally concentrates on the scalar spectrum
and assumes a perfectly smooth Einstein\,-\,de Sitter (EDS) angular\,-
\,diameter distance, \footnote{The angular\,-\,diameter distance  is the
natural extension of distance to cosmological situations, and is often called
the observer area\,-\,distance. It defines a distance equivalent to that
obtained using the apparent  luminosity of a source, see e.g.
\cite{bi:wein73,bi:elli71}.} which  neglects gravitational lensing and other
observational effects which are  unavoidable in a clumpy universe such as our
own. The influence of  gravitational lensing directly on the amplitude of the
CMBR temperature  anisotropies has been investigated, with contradictory
results, over the last  few years \cite{bi:Linder88,bi:linder90,
bi:Sasaki89,bi:Cole89,bi:Tomita89,bi:cayon93} by calculating the change to
angular functions such as the CMBR correlation function. However the impact
due to  gravitational lensing on the estimated {\it spatial} scalar power
spectrum  at decoupling  has so far been neglected. In this paper we address
this problem  by using a more suitable angular\,-\,diameter distance for our
universe than  the smooth EDS relation.

The angular\,-\,diameter distance lies at the heart of observational cosmology
since it is used, among other things, to convert observed angles to estimated
length scales on the surface of last scattering. We find that the assumption
of the EDS distance relation is unjustified in most cases, given the
observational accuracy available presently and conjectured for experiments in
the future.  We illustrate the change to the spatial power spectrum derived
from observations using the simplest model of an angular\,-\,diameter
distance which allows for inhomogeneities \cite{bi:Dyer73} as an example.
Here we assume the Einstein field equations, and an  ``almost\,-\,Copernican"
principle which, when combined with the almost\,-\,isotropy of the CMBR,
implies that the universe is almost Friedmann\,-\,Robertson\,-\,Walker (FRW)
on large enough scales \footnote{The proof that almost\,-\,isotropic
observations implies almost\,-\,FLRW behaviour has recently been rigorously
discussed in \cite{bi:SME94} (see also \cite{bi:BF91}).}. Thus we assume that
the large\,-\,scale dynamics of the universe can be described by (generalized)
FRW parameters.
%We do not however exclude a nonlinear density power spectrum at small scales,
%although we do not treat it quantitatively
%here because of the existance of caustics.
%%%%%%%%%%%%%%%%%%%%%%%%%%%%%%%%%%%%%%%%%%%%%%%%%%%%%%%%%%%%
\section{Measuring  cosmological distances} \label{sec:shrink}
%%%%%%%%%%%%%%%%%%%%%%%%%%%%%%%%%%%%%%%%%%%%%%%%%%%%%%%%%%%%%
In any universe, cosmological distances are measured using the
angular\,-\,diameter distance relation. In an inhomogeneous universe this will
differ in almost all cases from the corresponding relation in a smooth FRW
model \cite{bi:Linder88,bi:Dyer73,bi:Schneider92}. In a universe which
undergoes a transition to homogeneity on large scales,  such as may occur
with our own, the  angular\,-\,diameter distance may coincide exactly with
that of a FRW model when  averaged over these scales \cite{bi:Weinberg76}.
However, this depends critically on the areas of the wavefronts at the same
redshift being the same in the real universe as in the FRW model. The
existance of directions in the sky for which the wavefront  is multiply
sheeted due to gravitational lensing \footnote{Multiple imaging  causes the
wavefront to develop catastrophes and become multi\,-\,sheeted
\cite{bi:Schneider92}.} shows that this cannot be assumed, and implies that
the geometry of wavefront averaging associated with CMBR instrument beam
widths, is subtle. Since the angular\,-\,diameter distance is implicitly  used
to estimate the primordial density fluctuation spectrum from the observed
CMBR angular temperature variations across the sky, it is vital  that one uses
as accurate an angular\,-\,diameter distance as  possible.

In general the relation between a scale $l$ at redshift $z$, and the
angle $\theta_m$, it subtends when observed will depend on $\theta_m$,
the direction of observation  (represented by a 2\,-\,D vector $\bf{n}$) and
implicitly on the angular scales over which observations are averaged,
$\theta_r$, \footnote{In the case of the CMBR $\theta_r$ is the resolution of
the instrument. Detail smaller than this scale is lost.} viz:
\begin{equation}
l(z) = r(z,{\bf n},\theta_r,\theta_m) \theta_m\;.
\label{eq:one}
\end{equation}
Here $r(z,{\bf n},\theta_r,\theta_m)$ is the angular\,-\,diameter distance in
a general universe which will be anisotropic due to the shearing effects on
the ray bundle. We expect that this anisotropy will tend to zero (the
background FLW model) as the averaging scale increases, effectively  making
the angular\,-\,diameter distance isotropic in the limit of large averaging
scale. A fairly good approximation for the large\,-\,angle (e.g. {\it COBE})
experiments is that the shear is unimportant.  One can, however show that
small amounts of shear increase the convergence of the ray bundle and thus
decrease the angular\,-\,diameter
distance \cite{bi:Schneider92,bi:Weinberg76}. Larger amounts of shear
cause  conjugate points \cite{bi:Seitz93} to form in the past light cone which
leads  to image parity reversal and creation  of multiple images due to
self\,-\,intersection of the past null cone.

With this in mind and for illustrative purposes, we neglect the shear
\cite{bi:Dyer73} as a first approximation. It should be borne in mind that at
small angular scales this assumption is almost certainly unjustified.

The observed angular\,-\,diameter distance in the absence of shear becomes
locally isotropic  and can be characterized by the redshift only, so $r=r(z)$,
where
we will deal with the simpler conceptual case of an ideal experiment with
perfect
resolution, $\theta_r, \theta_m \rightarrow 0$  since the complications
of averaging are highly non\,-\,trivial. For $\Omega\leq 1$, all
inhomogeneities on
the surface of last scattering appear smaller than they would in a FRW model
and
hence when projected back onto the surface of last scattering  using an EDS
angular\,-\,diameter distance, appear much smaller than when estimated using
the
EDS model, all scales  being shrunk by the same amount  relative to this model
(hence the term ``shrinking"
\cite{bi:Linder88d})
\footnote{This implies that the fractal dimension of the intersection of the
last scattering surface and our past light cone, remains constant
\cite{bi:Bassett94c}.}, and therefore we expect the emphasis in the power
spectrum to shift to larger wavelengths (smaller $k$) due to the scaling
properties of the Fourier transform (the opposite occurs when $\Omega > 1$,
i.e. ``magnification'').

In general the observed CMBR will be a patchwork of regions with different
shrinking  factors, since we observe through a underdensities and
overdensities, so that a circular CMBR anisotropy will be seen to be dilated
uniformly  only if it is small and has an angular size close to the resolution
of the  instrument. Larger anisotropies will have different parts deformed by
different amounts, leading to a non\,-\,conformal mapping from the surface of
last scattering (or the spatial hypersurface the anisotropies were produced in)
onto the celestial sphere.

The best way to examine the competing effects of shrinking and magnification is
probably to define a spatial correlation function, in analogy with its angular
counterpart. We may define this on the spatial hypersurface where the dominant
contribution to the CMBR anisotropy on the angular scale of interest is
generated (for example, in an open universe, the large\,-\,angle ($>
10^{\circ}$)  anisotropies may be caused by low redshift structures
\cite{bi:KS94}, while at  small ($\approx 1'$) angles, the anisotropies are
due to Doppler effects in  the surface of last scattering).  One way to define
this is:
\begin{equation}
C(L,z) = \langle \frac{\delta T}{T}({\bf x + L})~
\frac{\delta T}{T}({\bf x}) \rangle
\end{equation}
where the vector ${\bf L}$ is replaced by a length $L$ on assuming ergodicity
and the difference operator is  between the point-, and average values to
ensure gauge-invariance. It is here that the difference between this work and
previous work can be seen. Although the effects of lensing on angular variables
has been examined, the change to the corresponding spatial functions has not.
The mapping from angular to spatial is usually carried
out using plane waves, if $\Omega = 1$. If $\Omega < 1$ then  this cannot be
done and the correct expansion eigenfunctions must be used \cite{bi:Wilson83}.
Thus even if the angular function is unchanged, the spatial function is
altered. Any change of geometry (from the flat, EDS one) will cause such a
modification of the spatial functions, and since lensing is a local modulation
of geometry, we expect our effect to be important, particularly on medium to
small scales. On large scales, we expect significant changes only if $\Omega_0
\not 1$, unless the universe does not make a transition to  homogeneity as in
the fractal models of structure.

In this paper we are concerned with the magnitude of the shrinking effect
relative to a EDS model, so we follow Dyer and Roeder (1973) and Schneider
{\it et al.} (1992) \cite{bi:Dyer73,bi:Schneider92} by introducing a
smoothness parameter $\alpha$ such that the matter in the universe is
described by two components; a smooth dust background of average density
$\alpha\Omega_d$ and a proportion $[1 - \alpha]\Omega_d$ in compact clumps,
which under the assumption of negligible shear, implies that they do not
affect the angular\,-\,diameter distance. We then define the ratio:
\begin{equation}
\gamma(\alpha,\Omega,z) = \frac{r_{\alpha,\Omega}(z)}{r_{1,1}(z)}\;,
\end{equation}
where $r_{\alpha,\Omega}(z)$ is the angular\,-\,diameter distance for a
universe with smoothness parameter $\alpha(z)$ and density parameter $\Omega$
at redshift$z$, (so $r_{1,1}(z)$ is the angular\,-\,diameter distance for a
pure FRW model with density parameter $\Omega = 1$). In particular we note
that in the FRW case the following relation exists between arbitrary $\Omega$
and $\Omega=1$ universes \cite{bi:Schneider92}:
\begin{eqnarray}
\gamma(1,\Omega,z) &=&
\frac{\Omega z - (2 - \Omega)
(\sqrt{\Omega z + 1} - 1)}{\Omega^2 (z + 1  - \sqrt{z + 1})} \\
&>& 1 ~~~~~\mbox{if} ~~~ \Omega < 1\;.
\end{eqnarray}
Thus we see that shrinking relative to the EDS model occurs even in
pure FRW models, but is enhanced by including
inhomogeneities. It follows that even in a smooth universe, the power spectrum
estimated from observations will be altered if $\Omega\not= 1$.

{}From equations (1) and (2) we see that a linear scale $l$ on the surface of
last scattering, subtending an angle $\theta_{EdS}$ in a smooth EDS universe,
actually subtends an angle:
\begin{equation}
\theta = \frac{1}{\gamma(\alpha,\Omega,z)} \theta_{EdS}\;.
\end{equation}
For example the angular size of the Hubble radius at decoupling, usually
quoted to be at an angle of about  $1^{\circ}$, actually subtends an angle
$\approx
1^{\circ} /\gamma$ where $\gamma$ defines the shrinking factor at the redshift
$z_{SLS} \approx 1000$,  corresponding to the surface of last scattering.

We include for generality a radiation energy density term $\Omega_r$, which
may be important in low $\Omega$ universes which are radiation dominated
at decoupling. Then the equation for the angular\,-\,diameter
distance is obtained from the transport equations for the optical scalars,
on neglecting the shear (see Appendix B):
\begin{equation}
P(z) \ddot{r} + Q(z) \dot{r} +
\left[\frac{3}{2}\Omega_d\alpha + 2(1 + z)\Omega_r\right] r = 0\;,
\label{eq:two}
\end{equation}
\begin{equation}
P(z) = (1 + z)(1 + \Omega _d z + z(2 +z)\Omega_r) \;,
\end{equation}
\begin{equation}
Q(z) = (3 + \frac{\Omega_d}{2} + \frac{7\Omega_d z}{2} + \Omega_r +
8\Omega_r z + 4\Omega
a_r z^2)\,.
\end{equation}
The corresponding initial conditions are ~\cite{bi:Seitz93}:
\begin{equation}
r(z=0) = 0\;, ~~~~~~\dot{r}(z = 0) = 1\;.
\end{equation}
This equation gives distances in units of $\frac{c}{H_0}$ and reduces to
the usual Dyer\,-\,Roeder equation when $\Omega_r = 0$
\footnote{The equation has the form of a general Mathieu equation with
parametric ``frequency'' and ``damping'' dependencies. Note that the ansatz
$r(z) = exp(-\case{1}/{2} \int \frac{Q(\eta)}{P(\eta)} d\eta) \chi(z)$  will
eliminate the $\dot{r}$ term.}. If in addition $\alpha(z) = \alpha_0$, a
constant, this equation can be converted into the hypergeometric equation
\cite{bi:Dyer73} or the Legendre differential equation \cite{bi:Seitz93}, so
that solutions exist for all $\alpha_0$ and $\Omega$. Finally Linder (1988)
has obtained the solutions for a large number of different limiting cases
\cite{bi:Linder88}.

In this paper we concentrate mainly on the $\Omega \approx 1$ case since
technical details regarding the redshift dependence of the anisotropy
generation and the definition of the power spectrum in other FRW models may
mask the underlying simplicity of the effect we are discussing.

For the numerical work \footnote{We integrated eq.(\ref{eq:two}) using a
Runge\,-\,Kutta\,-\,Merson method, which is ${\cal O}(h^5)$ and gives
an estimate of the local error.},
we choose a smoothness parameter of the form:
\begin{equation}
\alpha(z) =  1 - (1 - M_*) \left [ \frac{1 - \alpha_5}{1 - M_*} \right ]^
{\frac{\textstyle{z}}{5}}\;,
\end{equation}
where $M_* \equiv \alpha(0)$ is the present proportion of mass in
smoothly distributed form and $\alpha_5 \equiv \alpha(5)$ is the same
quantity at a redshift of $5$, which corresponds to the present outer limit of
observed  quasars. We parametrise $\alpha_5$ by an
evolution index, $p, (p >1)$, which is constant for each simulation, so that:
\begin{equation} \alpha_5 \equiv \alpha(z=5) =  \frac{M_* + p - 1}{p}\;,
\end{equation}
with more rapid evolution into clumps simulated by using smaller $p$.
The neutral baryonic component in the observable universe is almost entirely
in clumpy form today and even at $z = 5$ there appears to be very little
neutral gas in the intergalactic medium \cite{bi:barkana94} (using
the standard interpretation of the Gunn\,-\,Peterson test).

The two parameters $(M_*,p)$ are more difficult to estimate in the case of
dark matter, which is a partial restatement of the bias problem. Only a fairly
small proportion of the dark matter is expected to be in compact form today
since the inhomogeneities are still approximately linear on scales larger than
about $10 h^{-1} Mpc$. In particular it should be noted that $\alpha(z)$
depends on the scale one is  averaging over to obtain the ``clumps'', with the
limiting behaviour that $\alpha(z) \rightarrow 1$ when averaged over the whole
sky. Thus the shrinking  factor is strictly speaking dependent on
observational scale even when shear  is neglected, making the impact on the
estimated power spectrum even more intricate.

Because of this uncertainty, and for generality, we plot our results for a
wide range of the two parameters $M_*$ and $p$ (see figures 2\,-\,4 and table
1 below).
\begin{table}
\def\tablerule{\noalign{\hrule}}
$$\vbox{\tabskip=0pt \offinterlineskip
\halign to 311pt{\strut#&\vrule#\tabskip=1em plus 2em&
&#\hfil&\vrule#&#\hfil&\vrule#&#\hfil&\vrule#&#\hfil&\vrule#&#
\hfil&\vrule#&#\hfil&\vrule#
\tabskip=0pt\cr\tablerule
\omit&height2pt&&&&&&&&&&\cr
&&$M_*$&&$p$&&$\gamma$&&$\gamma^{6+n}~~(n=1.5)$
&&$\gamma^{6+n}~~(n=1)$&\cr
\omit&height2pt&&&&&&&&&&\cr\tablerule
&&$0.95$&&2.0&&$1.08$&&$1.78$&&$1.71$&\cr\tablerule
&&$0.9$&&$1.5$&&$1.26$&&$5.66$&&$5.04$&\cr\tablerule
&&$0.8$&&$2$&&$1.40$&&$12.47$&&$10.54$&\cr\tablerule}}$$
\caption{Shrinking factors, $\gamma$, for various evolutionary parameters (n
is the spectral index). See below for discussion of $M_*,p$.}
\end{table}
For $\Omega = 1$ we plot the range of $\gamma\,-\,$factors as a
function  of $M_* \equiv \alpha(z = 0)$, the smoothness parameter today. We
take as the  smallest reasonable  $\gamma\,-\,$factor, the value
$\gamma_{min} = 1.08$.  Allowing $\Omega < 1$ will drastically increase the
$\gamma\,-\,$factor, so  that $\gamma_{min}$ provides a rigorous lower bound
for this effect in models  with $\Omega \leq 1$ in the approximation of
neglecting the shear \footnote{ Although small amounts of shear may decrease
the $\gamma-$factor, the effect  on the CMBR fluctuations will still be
non\,-\,trivial because the  angular\,-\,diameter distance, although smaller,
will become anisotropic, so  that shrinking will be a direction dependent
phenomenon.}

There is an interesting question which is also of practical importance:
what is the asymptotic behaviour ($z \rightarrow \infty$) of
$\gamma(\alpha,\Omega,z)$ ? This is important for stability reasons since the
surface of last scattering is ill\,-\,defined and has finite thickness, with
redshift
estimates ranging between $z_{SLS} = 1000 - 1500$. In figure 4 we plot
$\gamma(z)$
against redshift for various evolution indices and clumpiness parameters. We
find that $\gamma(z)$ monotonically increases with redshift. This result
implies that our estimates of the shrinking effect will be slightly increased
if we assume the surface of last scattering lies at $z_{SLS} = 1500$ instead
of $1000$. For very large redshifts the shrinking factor may approach a non\,-
\,zero constant. It is easy, though fairly tedious, to show that a constant
$\gamma$ is a solution of the governing evolution equation in the asymptotic
region  $z \rightarrow \infty$ and we conjecture that this constant should be
determined by the low\,-\,redshift behaviour of $\alpha(z)$, since
asymptotically $\alpha(z) \rightarrow 1$ for generic perturbed FRW models.
%%%%%%%%%%%%%%%%%%%%%%%%%%%%%%%%%%%%%%%%%%%%%%%%%%%%%%%%%
\section{The new power spectrum} \label{sec:power}
%%%%%%%%%%%%%%%%%%%%%%%%%%%%%%%%%%%%%%%%%%%%%%%%%%%%%%%%%
Let us now turn to an explicit calculation of how the power spectrum changes
when we include the effects of shrinking. By scaling all fluctuations on the
surface of last scattering by a factor  $\gamma \equiv
\gamma(\alpha,\Omega,z=1000)$ and then taking its Fourier transform we find
that the new power spectrum $|\hat{\delta}_k|^2 \equiv |\hat{\delta}(k)|^2 $
is related to the old one, $|\delta (k)|^2$ by (see Appendix A for the
derivation):
\begin{equation}
\hat{\delta}(k) = \gamma^{3} \delta(\gamma k) \;.
\label{eq:khat}
\end{equation}

In the case of a power\,-\,law spectrum which is expected on
large and medium scales at decoupling, shrinking changes our estimation of it
by increasing the amplitude via:
\begin{equation}
|\delta (k)|^2 = Ak^n  \Rightarrow |\hat{\delta}(k)|^2 = (A \gamma^{6 + n})
k^n\;,
\end{equation}
where $A$ is the normalization constant. The spectral index is
unchanged \footnote{The invariance of $n$ under shrinking is only true in the
approximations made in this paper. $n$  will be a function of $\gamma$ when
the scale\,-\.dependence of shrinking due to shear and averaging is included.}
in the case of a power\,-\,law spectrum (since power laws are scale invariant).

Thus in the power\,-\,law regime the true amplitude of the
spectrum at decoupling is larger by the factor $\gamma^{6 + n}$ than that
found from observations assuming an EDS model.

The spectrum at decoupling is unlikely to be a power\,-\,law at all scales.
Even for theories that predict a power\,-\,law primordial spectrum such as
many inflationary and cosmic string models, at small scales the spectrum at
decoupling will have had this symmetry broken by gravitational and other
physical processes.

One other point should be noted in regard to determining the power spectrum on
the surface of last scattering. In our approximation, the angular\,-\,
diameter distance is assumed to depend only on redshift. However, to ensure
that the results are gauge\,-\,invariant, one must define the surface of last
scattering in a gauge\,-\,invariant way too. In particular one must specify how
this perturbed hypersurface lies in spacetime relative to the background model
(with its smooth surface of last scattering). A result of this which we
mention is that the redshift of the real surface of last scattering will not
be the same in different directions, and hence the angular\,-\, diameter
distance will vary across it, causing {\it anisotropic} variations in the
shrinking, $\gamma$. This will induce  changes into the estimated spatial
power spectrum, which cannot be modelled using eq.(11) (see  appendix B).
However, since these redshift fluctuations are expected to be  relatively
small, we also expect the variation in $\gamma$ to be small,
particularly at high redshift (see figure 4).

Considering only the effects of gravitation and free streaming
[ignoring photon diffusion, Sakharov oscillations and the integrated Sachs-
Wolfe effect \cite{bi:Kodama86,bi:Hu94}] on an initially power law spectrum,
we find that it increases with $k$ to a maximum and then turns around and
decreases for $k > k_{eq} \approx \frac{\Omega^2}{2}$, where $k_{eq}$ is the
wave number corresponding to the Hubble scale  in Mpc$^{-1}$, at the change
over from radiation to matter domination. The theoretical functional form of
the power spectrum for $k \gg k_{eq}$ is predicted to be
\cite{bi:Padmanabhan93}:
\begin{equation}
|\delta(k)|^2 = B\frac{\ln^2(k/k_{FS})}{k^3} ~~~~~~~~ k \gg k_{eq}\;,
\label{eq:three}
\end{equation}
where $B$ is a normalization constant determined by matching to the spectrum at
large scales and $k_{FS}$ describes the smallest perturbations which are not
washed out due to the free\,-\,streaming of the particles forming the dark
matter. If this is the true form of the spectrum at small scales, then we will
observe something different due to the shrinking effects of the
inhomogeneities (not to mention the effects of strong lensing through caustic
formation.)

The power spectrum at small scales, \footnote{Where the shrinking process is
unaffected by large\,-\,scale homogeneity arguments.} {\em assuming}
it takes the form given in equation (\ref{eq:three}), is obtained by
``inverting'' the process followed at large\,-\,scales, since we now wish to
predict the observed spectrum, and not the true spectrum. By using our formula
for the new spectrum, equation (\ref{eq:khat}), one obtains:
\begin{eqnarray}
|\hat{\delta}(k)|^2&=&B \gamma^{-6 + 3}k^{-3}
\left[\ln(k/(\gamma k_{FS})^2\right]\nonumber\\
&=&\gamma^{-6+3} \left\{ |\delta(k)|^2  -  \frac{B}{k^3} \ln \gamma
\left[ \ln(\frac{k^2}{(k_{FS}^2 \gamma)}) \right] \right\}\;,
\label{eq:small}
\end{eqnarray}
and hence (for $\gamma > 1$)~~ $|\hat{\delta}(k)|^2<|\delta (k)|^2$ for this
$k$-range, which shows that shrinking will lead to observations which
over\,-\,estimate the true spectrum and hence lies below the spectrum derived
from observations ignoring shrinking. This may be important for CMBR
experiments at scales less than a degree such as the OVRO \cite{bi:Readhead89}
and South Pole 89 \cite{bi:Meinhold91} experiments. It is clear that shrinking
causes richer effects at small scales than in the large angle scale\,-\,free
case, and we conjecture that it may have a role to play in the apparent
inconsistancy of the small\,-\,scale experiments \cite{bi:Wright94} with {\it
COBE}.

If one solves the perturbed kinetic theory problem for the coupled photon-
baryon fluid before decoupling one gets the Doppler peaks, or Sakharov
oscillations. The effects of lensing, particularly on the position of these
Doppler peaks is important, since it has been proposed \cite{bi:KSS94} that
the position of the first Doppler peak can be used to determine the value of
$\Omega$. Since all of our results show that the angular power spectrum is
shifted relative to the one estimated without considering lensing, we claim
that shrinking will change the position of the Doppler peaks.
In particular, the position of first Doppler peak is determined by the angular
size of the Hubble scale at decoupling, which is determined by (see eq. 6,17)
the geometry and dependent on $\gamma$.  Hence, unless lensing effects are
included, the value of $\Omega$ estimated using small\,-\,angle experiments in
the future, could be wrong. Detailed work including the effects of caustics
should be included when examining the very\,-\,small scale experiments planned
for the future.

Until now we have assumed a globally flat universe ($\Omega=1$). Here we will
briefly discuss shrinking in an open model. This is important since most
observations indicate that $\Omega < 1$ \cite{bi:CE94,bi:peeb93}. In this case
the average energy density is significantly  smaller and hence the angular\,-\,
diameter distance at a given $z$ is much larger in these models.  It should be
emphasised that this is completely independent of the  effects of lensing.

If the geometry of the universe is open ($\Omega < 1$) then the effects are
much more drastic, primarily because the change to the eigenfunctions is
large. This can lead to $\gamma \approx 10$, which means a very large change
to spatial functions estimated at redshifts of $1000$. Naive application of
eq. (12) to the power spectrum suggests a correspondingly large change to the
normalisation of the {\it COBE} spectrum in an open universe.
However, the definition of power spectrum differs from the flat geometry case,
and it is no longer so useful to calculate the new spatial power spectrum at a
single redshift, since in low\,-\,$\Omega$ models the dominant cause of
large\,-\,angle {\it adiabatic} anisotropies is due to a combination of the
standard Sachs\,-\,Wolfe term at the surface of last scattering, together with
an integrated  effect gathered along the past null cone, but particularly at
low redshifts, $z\approx \frac{1}{\Omega}$, when the universe is curvature
dominated \cite{bi:KS94}. Hence it is difficult to say what the physical
significance is.

The effect of lensing in open models is similarly more complicated, since the
anisotropies generated by the Sachs\,-\,Wolfe effect will be lensed all the
way from last scattering at $z_{SLS} \simeq 1000$. Similarly, any nonadiabatic
perturbations, causing intrinsic temperature fluctuations in the surface of
last scattering, can be substantially lensed together with the Doppler peak
anisotropies. The anisotropies generated by the  integrated Sachs-Wolfe
effect on the other hand, are predominantly produced at $z\approx
\frac{1}{\Omega}$ and hence have little time to be lensed.
%%%%%%%%%%%%%%%%%%%%%%%%%%%%%%%%%%%%%%%%%%%%%%%%%%%%%%%%%%%%
\section{The new mulitpole moments} \label{sec:quad}
%%%%%%%%%%%%%%%%%%%%%%%%%%%%%%%%%%%%%%%%%%%%%%%%%%%%%%%%%%%%
The standard technique in CMBR experiments is to perform a spherical harmonic
decomposition of the temperature fluctuations: \begin{equation} \frac{\Delta
T}{T} = \sum_{\ell,m} a_{\ell,m} Y_{\ell,m}(\theta,\phi)\;. \end{equation} For
large  $\ell$, we have the relation \cite{bi:Liddle93}: \begin{equation}
\ell_d \approx \frac{57^{\circ}}{\theta}\;, \end{equation} between the
characteristic angular size $\theta$ of an anisotropy and the order $\ell_d$,
of the multipole which dominates its angular decomposition. Now if we could
observe an inhomogeneity on the surface of last scattering through a smooth
FRW universe, the angle subtended would be larger than in the true universe,
by a factor $\gamma$ (see figure 1).

Since the analysis of {\it COBE} data uses the observed angles ($10^{\circ}$
and
larger) but assumes an EDS model, the dominant order, $\ell_d$,
is larger than it is in a clumpy universe. Thus power flows to the
lower order harmonics when inhomogeneity is taken into account.
An important result, as shown below, of this low\,-\,order $\ell$
enhancement, is that it changes the {\it rms} temperature fluctuation  and the
magnitude of the quadrupole, $Q$ which are measured by the
{\it COBE} ~\cite{bi:Smoot92}.

Now the formula \cite{bi:Padmanabhan93} relating coefficients
$C_{\ell}\equiv\langle |a_{\ell,m}|^2\rangle$ to the spatial power spectrum
is a linear functional of $|\delta(k)|^2$, given (if $\Omega = 1$) by:
\begin{equation}
C_{\ell} = \frac{H_0^4}{2\pi} \int_0^{\infty} \frac{|\delta(k)|^2}{k^2}
|j_{\ell}
(k \eta)|^2 dk\;,
\end{equation}
where $j_{\ell}$ is the spherical Bessel functions of order $\ell$ and
$\eta$ is the difference in conformal time between the present and
decoupling epochs. If $\Omega \not = 1$ then the correct functions instead
of $j_{\ell}$ must be used, encoding the differences discussed at the
end of the last section. An additional effect of inhomogeneity is to
perturb these functions, which are only exact for constant curvature
spatial sections. These backreaction effects may be quite large
given the large arguments, $k\eta$, for the surface of last scattering.

Since {\it COBE} looks at large scales, where the spectrum is likely to be a
power\,-\,law, each of the $C_{\ell}$ is simply modified by the
proportionality factor $\gamma^{6 + n}$ \footnote{Since $|\delta(k)|^2 \mapsto
\gamma^{6 + n} |\delta(k)|^2$ in the case of a power\,-\,law} when the
new spectrum is substituted for the old.
This affects both the {\it rms} and the quadrupole values for the temperature
fluctuations \footnote{Note that the ratio of the two is unchanged in our
approximation, although there will be a change when large\,-\,$\ell$
variations are included in the {\it rms} formula [equation (\ref{eq:small})].
Similarly we expect the bispectrum \cite{bi:Luo93} to be qualitatively
unchanged by weak lensing - it is still expected to be zero for a Gaussian
power spectrum. This is probably not true when caustics are included.}.

There is one subtlety that must be taken into account here: any instrument has
a finite angular resolution, $\theta_r$, which means that the final instrument
spectrum is the true signal convolved with the response function of the
instrument. This means that the {\it rms} and quadrupole (Q) temperature
fluctuations obey the modified equations \cite{bi:Padmanabhan93,bi:Hu94}:
\begin{equation}
\left (\frac{\Delta T}{T}\right)^2_{rms} = \frac{1}{4\pi} \sum_{\ell =
2}^{\infty}
(2\ell + 1) C_{\ell} \exp(-\frac{\ell(\ell + 1) \theta_r^2}{2})
\end{equation}
and
\begin{equation}
\left (\frac{\Delta T}{T}\right)^2_{Q} = \frac{5}{4\pi}C_2 \exp(-2\theta_r^2)
\end{equation}
where the instrument is characterized by $\theta_{FWHM}$, which with
$\theta_{FWHM} \approx 7^{\circ}$ for {\it COBE} DMR, gives $\theta_r \approx
3^{\circ}$.

%%%%%%%%%%%%%%%%%%%%%%%%%%%%%%%%%%%%%%%%%%%%%%%%%%%%%%%%%%%
\section{The issue of  averaging} \label{sec: disc}
%%%%%%%%%%%%%%%%%%%%%%%%%%%%%%%%%%%%%%%%%%%%%%%%%%%%%%%%%%%

Let us consider a large\,-\,angle experiment such as {\it COBE} DMR or
Tenerife. At these scales what effects will lensing have on the CMBR ?  If
$\Omega \not = 1$ then the effect will be dominated by the new background
spatial eigenfunctions. No longer can one expand in plane waves, but instead
one must use the correct eigenfunctions of the Laplacian on the background
geometry. If $\Omega < 1$ then there is exponential divergence of geodesics
\cite{bi:KS94} and a very large change to the angular\,-\,diameter distance.
The effect of lensing will be to cause modulations about the background
geometry angular\,-\,diameter distance. A narrow beam may travel through a
globally underdense  region or it might travel through a globally overdense
region, in which case it  experiences an average $\Omega$ less than or greater
than the global mean and hence there is shrinking or magnification.
Investigations over the years \cite{bi:Bertotti66,bi:FME94,bi:Dyer88} have
shown that generically,  shrinking affects most lines of sight, (i.e. most
lines of sight are demagnified) with a few lines of sight highly magnified.
This effect can be understood in the following way. Consider equi-density
contour surfaces in the spatial hypersurfaces. We expect the volume
inside an underdensity surface to be larger than that of the corresponding
overdensity surface, and hence to subtend a larger solid angle generically on
the sky at the same redshift. The distribution  functions are skewed relative
to each other. Thus an average line of sight is more likely to intersect an
underdensity than an overdensity, but when it does intersect an overdensity,
the magnification is much larger if the total luminosity is the same as the
{\it corresponding} FLRW universe \footnote{This last line sums up much of the
difficulty faced in the subject since it requires that we have a well-
defined averaging and fitting procedure.}

Do the two effects cancel out ? Consider the following argument, in an $\Omega
= 1$ universe,  where the large-angle anisotropies are produced at the surface
of last scattering via the Sachs-Wolfe effect. The discovery by COBE of
temperature differences on $60^{\circ}$ scales after each horn averages over
$10^{\circ}$ implies that there are density fluctuations at angular scales
larger than $10^{\circ}$ - this is the Grishchuk-Zel'dovich effect whereby
superhorizon modes at decoupling contribute to the anisotropy. Hence we expect
density fluctuations at all smaller scales. Thus after  averaging over
$10^{\circ}$, we still do not have constant density over the  sky. This
contrast only grows since $10^{\circ}$ today is well inside the  horizon,
where nonlinearities dominate. Thus we conjecture that homogeneity  has not
been reached and hence the effects may be important even at {\it COBE}
scales, {\it assuming} that the temperature fluctuations directly trace
density fluctuations as  in the $\Omega = 1$ Sachs-Wolfe case. If entropy
perturbations are important, or $\Omega < 1$, this may not be true
\cite{bi:KS94}. Certainly at small scales, it will be important, and hence
needs to be taken into account for experiments at these scales.

%%%%%%%%%%%%%%%%%%%%%%%%%%%%%%%%%%%%%%%%%%%%%%%%%%%%%%%%%%%
\section{Conclusions } \label{sec: concl}
%%%%%%%%%%%%%%%%%%%%%%%%%%%%%%%%%%%%%%%%%%%%%%%%%%%%%%%%%%%
In this paper we have asked the question of how inhomogeneity (or indeed a
$\Omega \not = 1$ geometry) will change estimates of spatial functions at
decoupling, in particular the mass fluctuation power spectrum. Here are
several approaches we could use in future to examine this effect: (1) One
could find the direct change to the angular\,-\,diameter distance due to
inhomogeneity. This is the approach adopted in this paper, although only in a
simplistic way. (2) One could examine the effect of changes to the
background geometry eigenfunctions used to convert from angular to spatial
functions. An alternative way of formulating this is in terns of the effect of
{\it backreactions of the inhomogeneity on the interpretation of
observations}.  It has already been shown \cite{bi:BF89,bi:BF91} that although
the backreactions on the geometry may be small, the effects on obervational
estimates can be large.

At small scales, where the effect may be most important, the results are more
interesting since the spectrum will not be power\,-\,law. An application is
to the complex Doppler peak distribution in the spectrum at small-scales.
It has been suggested that the position of the first Doppler peak is sensitive
to the value of $\Omega$, but insensitive to most other physical  variables,
and hence could be a test of the geometry of the universe. Here
we claim that shrinking (we use the name generically to denote the type of
effect and not necessarily just the underestimation found in our simple
approximation) will shift the position of the Doppler peak, as will
other facets of lensing which are particularly important at small scales.
%\cite{bi:CMS94}.

It is important to answer the question of how shrinking differs from other
investigations of lensing of the CMBR, which have invariably concentrated on
the change to the {\it angular} correlation function.  The standard method
uses perturbed geodesics on a flat background, a method in some way connected
to using perturbed eigenfunctions. However, previous authors have only
calculated the perturbation effects up the null cone, since they were only
interested in angular functions on the celestial sphere. However, to calculate
the change to spatial functions at a given redshift, as required for
comparison with structure formation theories, one cannot use smooth background
formulae. One must recalculate the effect when going down the null cone in
converting angles to distances by including the effects of inhomogeneity.

One of the features of modern observational cosmology is the emergence  of
clustering on scales larger  than can be explained by standard cold dark
matter (CDM) theory with an $n = 1$ primordial spectrum \cite{bi:Maddox}. This
can be aliviated by putting more power into small $k$ \,-\, i.e.  choosing $n <
1$. However this is not supported by the latest CMBR experiment results
\cite{bi:Wright94}, with {\it COBE} DMR giving $n = 1.46 ^{+0.41}_{-0.44}$
(although this includes the low quadrupole and is calculated using
non-orthonormal basis functions). Alternatively, one may
introduce a non-negligible gravitational wave spectrum which allows the scalar
spectral index to be less than one. Another way of looking at this problem is
that CDM, normalized to the {\it COBE} results, produces too much
inhomogeneity on small scales \cite{bi:Ostriker93}. Using eq. (\ref{eq:small})
we see that shrinking may aid CDM significantly, since small scale
inhomogeneity is reduced due to shrinking.

Shrinking, with the change it induces in the quadrupole  moments of the
density and gravitational wave spectra, will also change the inflationary
potential reconstructed from observations of the CMBR \cite{bi:Turner93} and
may affect conclusions regarding the thermal \footnote{A late reionisation of
the  IGM may mimic a gravitational wave component of the CMBR spectrum
\cite{bi:Crittenden93}. } history of the universe based on CMBR observations.

Although we considered a simple angular\,-\, diameter distance, which only
depended on redshift, we show (appendix B) that the scaling of the power
spectrum given by eq.(11) is valid for distance relations which depend on beam
width (averaging scale). As soon as the angular\,-\, diameter distance varies
with direction in the sky (such as when the anisotropies in the redshift of
the surface of last scattering are considered) or with the size of the
inhomogeneity, the formula is incomplete.

As with previous investigations on the effects of inhomogeneity and
gravitational lensing on the propagation of light, we find that our results
depend fairly sensitively on the amount of matter in non\,-\,linear, clumpy
form and on its evolutionary history  ~\cite{bi:Linder88,bi:Babul91}. We also
emphasize that the  main approximation made in this paper, namely the neglect
of the shear, may be unacceptable. However, pushing the surface of last
scattering closer to the big\,-\,bang ($z_{SLS} > 1000$), leads to an increase
in the amount the power spectrum at decoupling is suppressed. A period of
reionization at late times ($z \approx 40$) would only lead to a small
reduction of the effect (see fig. 4). More sophisticated analysis is required
to determine what influence shear has, the impact of caustics and the scale-
dependence of shrinking, but the results obtained  from this conceptually
simple effect, highlight some of the  pitfalls that abound in making use of
standard models. The same reason makes analysis of the  implications for
structure formation and inflationary models worthwhile.  We hope that the full
picture regarding the effects of gravitational lensing will become clearer,
for only then can one be truly confident  about the implications of
observations, particularly of the CMBR.
%%%%%%%%%%%%%%%%%%%%%%%%%%%%%%%%%%%%%%%%%%%%%%%%%%%%%%%%%%
\section*{ Acknowledements}
%%%%%%%%%%%%%%%%%%%%%%%%%%%%%%%%%%%%%%%%%%%%%%%%%%%%%%%%%%
The authors would like to thank Igor Barashenkov for
help at critical stages of writing. They would also like to thank the referee,
Uro$\breve{s}$ Seljak, Malcolm MacCallum and Nazeem Mustapha for comments on
the final draft. BACCB would like to thank Jeffrey Cloete for enlightening
discussions over the years.
%%%%%%%%%%%%%%%%%%%%%%%%%%%%%%%%%%%%%%%%%%%%%%%%%%%%%%%%
\appendix
%%%%%%%%%%%%%%%%%%%%%%%%%%%%%%%%%%%%%%%%%%%%%%%%%%%%%%%%%
\section{Proof of the validity of a varying $\alpha$}
%%%%%%%%%%%%%%%%%%%%%%%%%%%%%%%%%%%%%%%%%%%%%%%%%%%%%%%%%
Consider the energy evolution equation with a variable $\alpha$. We wish to
show
that a varying $\alpha$ is valid in the approximation we make. We start with
the equation of energy conservation in a dust FLRW universe:
\begin{equation}
\dot{\mu} + \mu \theta = 0
\end{equation}
Then if one substitutes $\mu \rightarrow \alpha(t) \mu$ and solves this
equation,
one finds that $\mu \propto \alpha^{-1} S^{-3}$, where $\theta = \frac{1}{3}
\frac{\dot{S}}{S}$. Thus the matter evolves unchanged. To see that equation
(7) remains the same under this transformation, let us start with the general
equation for the evolution of an infinitesimal area of a ray bundle under the
assumption of vanishing shear  to get: \cite{bi:Schneider92}
\begin{equation}
\frac{d^2 r}{d w^2} + 4\pi(1 + z)^2 \rho r = 0
\end{equation}
making the ansatz $\rho(z) = \alpha(z)(1 + z)^3$ where $\alpha(z)$
characterises the deviation of the ``intergalactic" density from the global
average, and then changing from affine parameter, $w$, to redshift, $z$ gives
eq.(5), without involving the density.

Notice that the spatial sections are still homgeneous and isotropic in our
approximation. The full equation involves a shear term, (see
\cite{bi:Schneider92}
p. 139, eq. (4.49)) which definitely will depend on
redshift, thus we may define a ``generalised" $\alpha(z)$ which mimics the
effects of shear - not something we have tried to do - but showing the
functional validity of the equation with non\,-\,constant $\alpha$.
%%%%%%%%%%%%%%%%%%%%%%%%%%%%%%%%%%%%%%%%%%%%%%%%%%%%%%%%%
\section{Derivation of the new power spectrum}
%%%%%%%%%%%%%%%%%%%%%%%%%%%%%%%%%%%%%%%%%%%%%%%%%%%%%%%%
Here we derive the new estimated power spectrum, scaled by shrinking, from
the one estimated assuming an EDS model. We assume as usual that
$ |\delta (k)|^2 = Ak^n $
where $A$ is the amplitude and $n$ the index of the spectrum. We want the new
spectrum $|\hat{\delta (k)}|^2$  after the scaling
\[ S(\bf{x}) \equiv \frac{\rho(\bf{x}) -
\overline{\rho}}{\overline{\rho}}~~ \mapsto
{}~~S(\frac{\bf{x}}{\gamma}) \]
where $S(\bf{x})$ is considered as a random function. $\rho(\bf{x}) $ is the
density field at decoupling, $\overline{\rho}$ is the average density and
$\gamma \equiv \gamma(\alpha,\Omega,z)$. Note that scaling all
inhomogeneity sizes by $\gamma$ (active transformation) is equivalent to
reducing the coordinate $\bf{x}$ by $\gamma$ to $\frac{\bf{x}}{\gamma}$
(passive transformation), as done above.

The Fourier transform of $S(\bf{x})$ in $3\,-\,$dimensional space is:
\footnote{ Strictly speaking $\delta(k)$ should depend on $\bf{k}$, but because
of the a
ssumed isotropy (ergodicity \cite{bi:Peacock91}) of the density
perturbations, we impose the
restriction that it depend only on $k$. This assumption is one of simplicity
imposed on the primoridial matter perturbations (Guassinity),
and will have to be reconsidered when caustics and the effects of shear are
included on estimates  of the spectrum because of the
anisotropy of the angular\,-\,diameter distance in that case.}
\begin{equation}
\delta (k)\equiv \delta_k = \int\frac{d^3 {\bf{x}}}{(2\pi)^3}
{e^{i \bf{k} \cdot \bf{x}}} S(\bf{x})
\end{equation}
and the new spectrum
\begin{equation}
\hat{\delta}(k) = \int\frac{d^3 {\bf{x}}}{(2\pi)^3 }
{e^{i \bf{k} \cdot \bf{x}}} S(\case{\bf{x}}/{\gamma})
\end{equation}
which  after  relabeling of the spatial variables  and some
rearrangement gives:
\begin{equation}
\hat{\delta}(k) = \gamma^{3} \delta(k\gamma) \hspace*{.5cm} ~
\end{equation}
so that
\begin{equation}
|\delta(k)|^2 = Ak^n  \Longrightarrow |\hat{\delta}(k)|^2 = (A \gamma^{6 + n})
k^n
\end{equation}

Let us now consider a  general angular\,-\, diameter distance, and the effect
it has. In this case $\gamma$ depends on redshift, direction, and beam width
(window function). How does the spectrum change ? Following through the above
calculations, we see that as long as the angular\,-\, diameter does't depend on
the
spatial coordinates in the surface of last scattering, then the result is the
same simple scaling by $\gamma^3$, where in this case
$\gamma = \gamma(z,\theta_r)$ However, as soon as the angular\,-\,diameter
distance depends on the direction in the sky, or the size of the inhomogeneity
observed, the modification is much more complex and depends explicitly on this
variation. Given the nature of CMBR experiments it thus seems a good
approximation, except when considering the effects of the angular variation
in redshift of the surface of last scattering. Thus the main problem lies in
finding
the best form for $\gamma$.
%%%%%%%%%%%%%%%%%%%%%%%%%%%%%%%%%%%%%%%%%%%%%%%%%%%%%%%%%%

\newpage

{\bf Figure 1:} A schematic diagram showing the essence of the shrinking
process: an inhomogeneity on the surface of last scattering subtends an
angle $\theta_m$ at the observer, $O$. When a EDS angular\,-\,diameter distance
is used to estimate the size of the inhomogeneity, the projected scale is
smaller than the true size.

\vspace{0.75cm}
{\bf Figure 2:} Plots of the angular\,-\,diameter distance, $r$, as a
function of the redshift. The three curves are for different present\,-\,day
smoothness parameters, $M_*$, with the smooth, Einstein\,-\,de Sitter model
corresponding to $M_* = 1$.

\vspace{0.75cm}
{\bf Figure 3:} The shrinking (magnification if $\gamma < 1$) factors as
a function of present-day smoothness parameter, $M_*$, for three different
evolution indices: $p = 1.5,2~ \&~ 4$. Large $p$ models have less matter in
non-linear form (although $\Omega = 1$ for all models) at the same epoch
than do small ones; hence
the increasing $\gamma$-factors with decreasing $p$. A realistic value for $p$
probably lies between $1.4$ and $3$, although this is dark-matter model
dependent.

\vspace{0.75cm}
{\bf Figure 4:} The behaviour of $\gamma$ is plotted as a function of
redshift out to $z_{SLS} = 1500$ for different $M_*$, the present\,-\,day
smoothness parameter. On this log plot it is clear that the
shrinking factor is increasing, although very slowly, between $z = 1000$
and $z = 1500$.

%\vspace{0.75cm}
%{\bf Figure 5:}
%The small scale angular spectrum with and without shrinking, normalized
%to unity at large scales. Shrinking causes a change in both the position and
%amplitude of the Doppler peaks.

\end{document}